\documentclass[reprint,superscriptaddress,amsmath,amssymb,aps,prl]{revtex4-2}
\usepackage{CJK}

\usepackage[english]{babel}
\usepackage{dcolumn}
\usepackage{amsmath}
\usepackage{graphicx}
\usepackage{graphicx}
\usepackage{enumitem}
\usepackage{lineno}
\usepackage{xcolor}
\usepackage{multirow}
\usepackage{textcomp}
\usepackage[separate-uncertainty=true]{siunitx}
\usepackage[colorlinks=true,linkcolor=blue,citecolor=blue,urlcolor=blue]{hyperref}

\newlist{aims}{enumerate}{1}
\setlist[aims,1]{
  label={Cut~\arabic*},
  leftmargin=*,
  align=left,
  labelsep=10mm,
  itemindent=\dimexpr\labelsep+\labelwidth+7pt\relax
}

\begin{document}

\newif\ifWordCount
\WordCountfalse

%\begin{CJK*}{GB}{} % Use default fonts from CJK (see below)
\title{High sensitivity characterization of an ultra-high purity NaI(Tl) crystal scintillator with the SABRE proof-of-principle detector}

\author{F.~Calaprice}
\email{spokesperson: frankc@princeton.edu}
\affiliation{Physics Department, Princeton University, Princeton, NJ 08544, USA}

\author{S.~Copello}
\affiliation{Dipartimento di Fisica, Universit{\`a} degli Studi di Genova and INFN Genova, Genova I-16146, Italy}

\author{I.~Dafinei}
\affiliation{INFN - Sezione di Roma, Roma I-00185, Italy}

\author{D.~D'Angelo}
\affiliation{INFN - Sezione di Milano, Milano I-20133, Italy}
\affiliation{Dipartimento di Fisica, Universit{\`a} degli Studi di Milano, Milano I-20133, Italy}

\author{G.~D'Imperio}
\affiliation{INFN - Sezione di Roma, Roma I-00185, Italy}

\author{G.~Di~Carlo}
\affiliation{INFN - Laboratori Nazionali del Gran Sasso, Assergi (L'Aquila) I-67100, Italy}

\author{M.~Diemoz}
\affiliation{INFN - Sezione di Roma, Roma I-00185, Italy}

\author{A.~Di~Giacinto}
\affiliation{INFN - Laboratori Nazionali del Gran Sasso, Assergi (L'Aquila) I-67100, Italy}

\author{A.~Di~Ludovico}
\affiliation{Physics Department, Princeton University, Princeton, NJ 08544, USA}

\author{A.~Ianni}
%\email{aldo.ianni@lngs.infn.it}
\affiliation{INFN - Laboratori Nazionali del Gran Sasso, Assergi (L'Aquila) I-67100, Italy}

\author{M.~Iannone}
\affiliation{INFN - Sezione di Roma, Roma I-00185, Italy}

\author{F.~Marchegiani}
\affiliation{INFN - Laboratori Nazionali del Gran Sasso, Assergi (L'Aquila) I-67100, Italy}

\author{A.~Mariani}
\email{corresponding author: ambra.mariani@gssi.it}
\affiliation{Gran Sasso Science Institute, L'Aquila I-67100, Italy}

\author{S.~Milana}
\affiliation{INFN - Sezione di Roma, Roma I-00185, Italy}

\author{S.~Nisi}
\affiliation{INFN - Laboratori Nazionali del Gran Sasso, Assergi (L'Aquila) I-67100, Italy}

\author{F.~Nuti}
\affiliation{School of Physics, The University of Melbourne, Melbourne, VIC 3010, Australia}

\author{D.~Orlandi}
\affiliation{INFN - Laboratori Nazionali del Gran Sasso, Assergi (L'Aquila) I-67100, Italy}

\author{V.~Pettinacci}
\affiliation{INFN - Sezione di Roma, Roma I-00185, Italy}

\author{L.~Pietrofaccia}
\affiliation{Physics Department, Princeton University, Princeton, NJ 08544, USA}

\author{S.~Rahatlou}
\affiliation{INFN - Sezione di Roma, Roma I-00185, Italy}
\affiliation{Dipartimento di Fisica, Sapienza Universit{\`a} di Roma, Roma I-00185, Italy}

\author{M.~Souza}
\affiliation{Physics Department, Princeton University, Princeton, NJ 08544, USA}

\author{B.~Suerfu}
\email{corresponding author: suerfu@alumni.princeton.edu}
\affiliation{University of California Berkeley, Department of Physics, Berkeley, CA 94720, USA}

\author{C.~Tomei}
\affiliation{INFN - Sezione di Roma, Roma I-00185, Italy}

\author{C.~Vignoli}
\affiliation{INFN - Laboratori Nazionali del Gran Sasso, Assergi (L'Aquila) I-67100, Italy}

\author{M.~Wada}
\affiliation{AstroCeNT, Nicolaus Copernicus Astronomical Center
of the Polish Academy of Sciences, Warsaw, Poland}

\author{A.~Zani}
\affiliation{INFN - Sezione di Milano, Milano I-20133, Italy}

\ifWordCount
\else
\begin{abstract}
We present new results on the radiopurity of a 3.4-kg NaI(Tl) crystal scintillator operated in the SABRE proof-of-principle detector setup. The amount of potassium contamination, determined by the direct counting of radioactive $^{40}$K, is found to be $2.2\pm1.5$~ppb, lowest ever achieved for NaI(Tl) crystals. With the active veto, the average background rate in the crystal in the 1-6~keV energy region-of-interest~(ROI) is $1.20\pm0.05$~counts/day/kg/keV, which is a breakthrough since the DAMA/LIBRA experiment. Our background model indicates that the rate is dominated by $^{210}$Pb and that about half of this contamination is located in the PTFE reflector. We discuss ongoing developments of the crystal manufacture aimed at the further reduction of the background, including data from purification by zone refining. A projected background rate lower than $\sim$0.2~counts/day/kg/keV in the ROI is within reach. These results represent a benchmark for the development of next-generation NaI(Tl) detector arrays for the direct detection of dark matter particles.

% Therefore, the background due to impurities in the bulk crystal is xxx~counts/day/kg/keV in the ROI, making it one of the most radiopure NaI(Tl) crystals ever made.

\end{abstract}

\maketitle

\fi

%\end{CJK*}

%\linenumbers

\newif\ifPRL
\PRLtrue

\newif\ifComment
\Commentfalse

\newcommand{\comment}[1]{
\ifComment
\textcolor{red}{#1}
\else
\fi
}

\ifPRL
\else
\section*{Introduction}
\fi

The existence of dark matter is widely accepted~\cite{dm-history}, yet the particle nature of dark matter is still an open fundamental question. Over the last 25~years, a series of experimental efforts have been made to search for the interaction of dark matter particles with ordinary matter in underground laboratories~\cite{dm-history}. However, almost all efforts yielded null results despite the impressive progress in background reduction techniques and detector technologies
~\cite{xenon1t-result, lux-results, pandaX-result, ds50-results, cresst-result, damic-results, supercdms-result, supercdms-result-erratum}.
Experimental efforts have mainly focused on the search for the so-called weakly-interacting massive particles~(WIMPs) introduced in 1985~\cite{witten-wimp}. At present, the best sensitivity has been obtained with a ton-scale liquid xenon time projection chamber: the spin-independent WIMP-nucleon scattering cross section is estimated to be less than $\sim$4$\times$$10^{-47}$~cm$^2$ for a 30-GeV WIMP~\cite{xenon1t-result}.

As an alternative way to probe particle dark matter, it was shown that the motion of the Earth around the Sun in the dark matter halo can induce an annual modulation in the dark matter interaction rate~\cite{drukier-modulation}. The modulation, which is of the order of a few \% in amplitude, has a specific phase that allows to discriminate against other non-modulating and modulating backgrounds. This approach has been exploited by the DAMA experiment~(short for DAMA/NaI and DAMA/LIBRA)~\cite{dama2020-summary}, which has been observing a clear annual modulation in its array of 250-kg extremely radiopure NaI(Tl) crystals. The annual modulation is consistent with the dark matter hypothesis, but in the standard WIMP framework, it is in tension with other more sensitive direct detection experiments~\cite{xenon1t-result,lux-results,pandaX-result,cresst-result,supercdms-result,ds50-lm-results}. Due to a potential target-dependence of WIMP-nucleon interactions, a verification using the same target material is indispensable. 

To test the DAMA annual modulation, ANAIS~\cite{anais-performance} and COSINE~\cite{cosine-initial-performance} are currently operating 112.5~kg and 106~kg of NaI(Tl) crystal scintillators, respectively. However, they could not provide a definite answer to the longstanding controversy due to their signal-to-noise ratio diminished by constant radioactive backgrounds several times higher than that observed in the DAMA crystals~\cite{anais2019-1st-result,cosine-result-modulation}. Large fractions of the backgrounds come from radioactive contaminants in the crystal, most notably $^{40}$K, $^{210}$Pb and $^3$H. This indicates that the definitive test of the annual modulation claim 
has to be addressed by next-generation experiments using NaI(Tl) crystals with radiopurity similar to or below the DAMA level, such as in the proposed SABRE experiment~\cite{sabre-pop}.

\ifPRL
\else
\section*{SABRE PoP and the veto performance}
\fi

\begin{figure}[ht]
  \centering
  \includegraphics[angle=0,width=\linewidth]
  {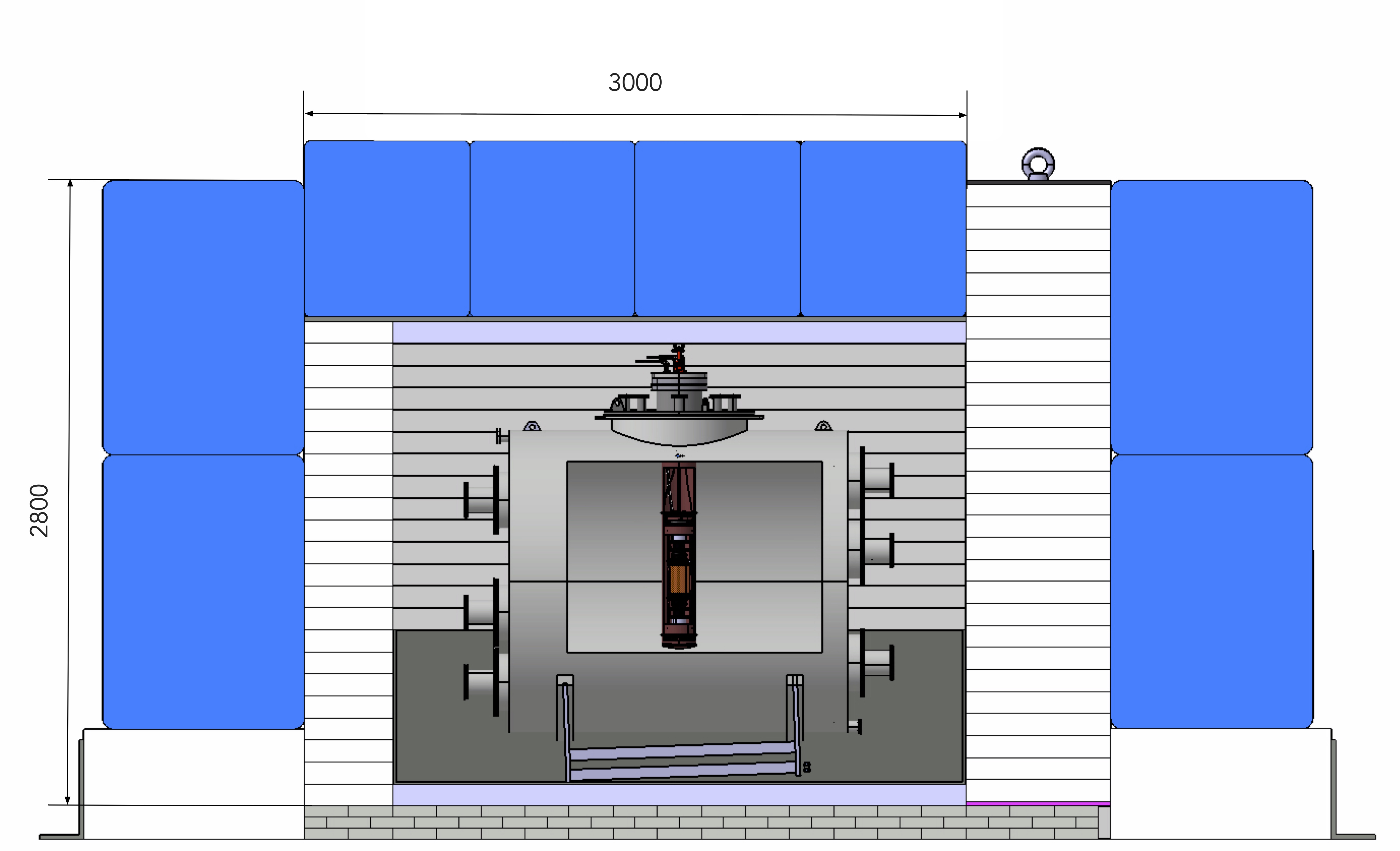}
  \caption{Technical drawing of the PoP system. From the outer to the inner region: water tanks~(blue), HDPE layers~(white), veto vessel~(light grey), and crystal enclosure~(brown).} \label{fig:1} 
\end{figure}

In this article, we report a detailed study of the background components of an ultra-high radiopurity crystal grown for the SABRE experiment. While preliminary measurements on the same crystal had been performed on the surface~\cite{nai033-surface} and underground inside a passive shielding~\cite{nai033-udg}, this study was carried out underground in the SABRE proof-of-principle~(PoP) liquid scintillator active veto shielding~(Fig.~\ref{fig:1}) in the Hall~C of Laboratori Nazionali del Gran Sasso~(LNGS), Italy~\cite{sabre-pop}.

This crystal---NaI-33---was grown using ultra-high purity NaI powder after a series of R\&D activities~\cite{phd-shields,phd-suerfu}. The preparation of the powder and the crucible were carried out at Princeton~\cite{phd-suerfu,nai033-surface} and the crystal growth was done at Radiation Monitoring Devices~(RMD) in Massachusetts using the vertical Bridgman method~\cite{nai033-surface,bridgman}. The crystal has a mass of 3.4~kg, and is wrapped with ten layers of PTFE tape~(about 1~mm) and optically coupled directly to 3-inch Hamamatsu R11065 photomultiplier tubes~(PMTs) that features high quantum efficiency and low radioactivity~\cite{nai033-surface}. The crystal-PMT assembly is sealed inside a 2-mm-thick, high-purity copper enclosure~\cite{nai033-udg}, whose inner volume is constantly flushed with nitrogen gas to purge moisture and radon. 
The enclosure is deployed vertically via a 2-mm-thick, 121-cm-long blind-end copper tube into a 1.3-m-diameter, 1.5-m-long stainless steel veto vessel~\cite{sabre-pop}. The veto vessel is filled with 2~tonnes of pseudocumene~(PC) liquid scintillator and $2.86$~g/L of 2,5-diphenyloxazole~(PPO) wavelength shifter, and is instrumented with ten Hamamatsu R5912 8-inch PMTs to serve as active veto to enhance background rejection~\cite{phd-shields,phd-suerfu}. The inner surface of the veto vessel is lined with reflective Lumirror foils to improve light collection~\cite{phd-suerfu}. The light yield of the active veto is measured to be $0.52\pm0.01$~photoelectrons/keV~(phe/keV) at 2615~keV using a $^{228}$Th calibration source. The veto vessel is further shielded from the cavern by an inner high-density polyethylene~(HDPE, 10~cm on the top and bottom and at least 40~cm on the sides) and an outer gamma shielding~($\sim$90~cm of water on the sides and top, and 15-cm of lead on the bottom)~\cite{sabre-pop}.
A PTFE tube is used to position wire-mount calibration sources next to the copper enclosure. 

%The liquid scintillator in the veto vessel, supplied by the adjacent Borexino plant~\cite{borexino}, is prepared by purifying by water extraction a concentrated solution of PC-PPO mixture and then mixing it with pure distilled PC. The final PPO concentration of $2.86\pm0.02$~g/L is determined by photospectrometric measurements on samples of the mixture. The light yield of the active veto was measured to be $0.52\pm0.01$~photoelectrons/keV~(phe/keV) at 2615~keV using a $^{228}$Th source placed near the center using the calibration system, consistent with a previous rough estimate~\cite{phd-suerfu}. By exploiting the coincidence between 583-keV and 2615-keV gamma rays, the energy resolution is determined to be $\sigma(E)/E = \sqrt{7.3/E}-20/E$, where E is expressed in keV.

The DAQ system consists of a CAEN~V1495 custom FPGA trigger and two CAEN~V1720 ADC boards~(12-bit, 250~MS/s)~\cite{phd-suerfu}. The data acquisition is triggered by the logical AND between the two PMTs coupled to the crystal with a 125-ns coincidence window  irrespective of the status of the veto detector~\cite{phd-suerfu}. Upon trigger, waveforms in the subsequent \SI{3.5}{\micro\second} window in all PMTs are digitized and read out by a dedicated DAQ software~\cite{sabre-daq}.

\ifPRL
\else
\section*{Background study}
\fi

\begin{figure}[ht] 
\centering
\includegraphics[angle=0,width=\linewidth]{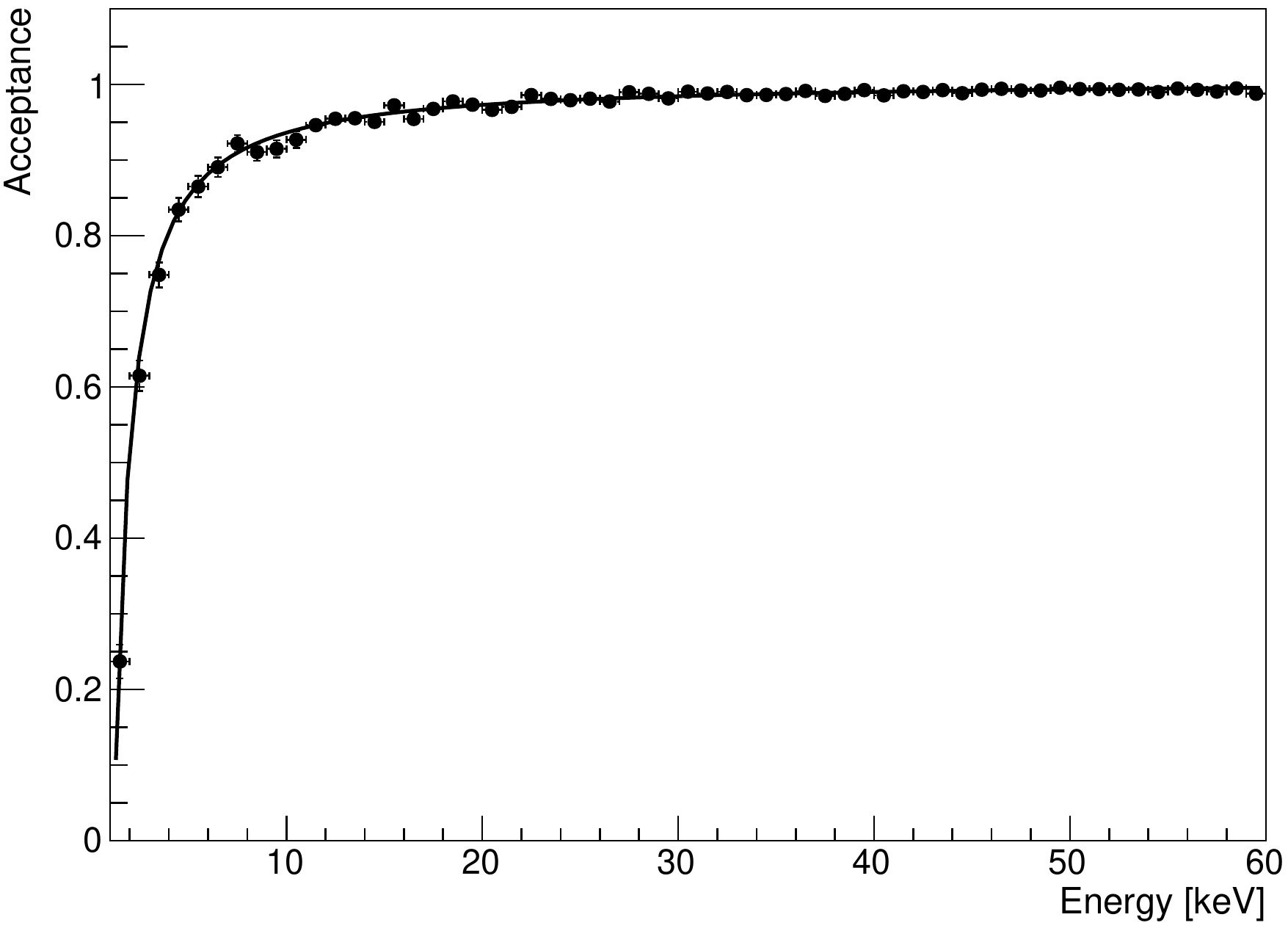}
\caption{Acceptance of crystal events after noise rejection cuts as a function of energy, evaluated using a $^{228}$Th calibration source.} \label{fig:2}
\end{figure}

\begin{figure}[ht]
  \centering
  \includegraphics[angle=0,width=\linewidth]
  {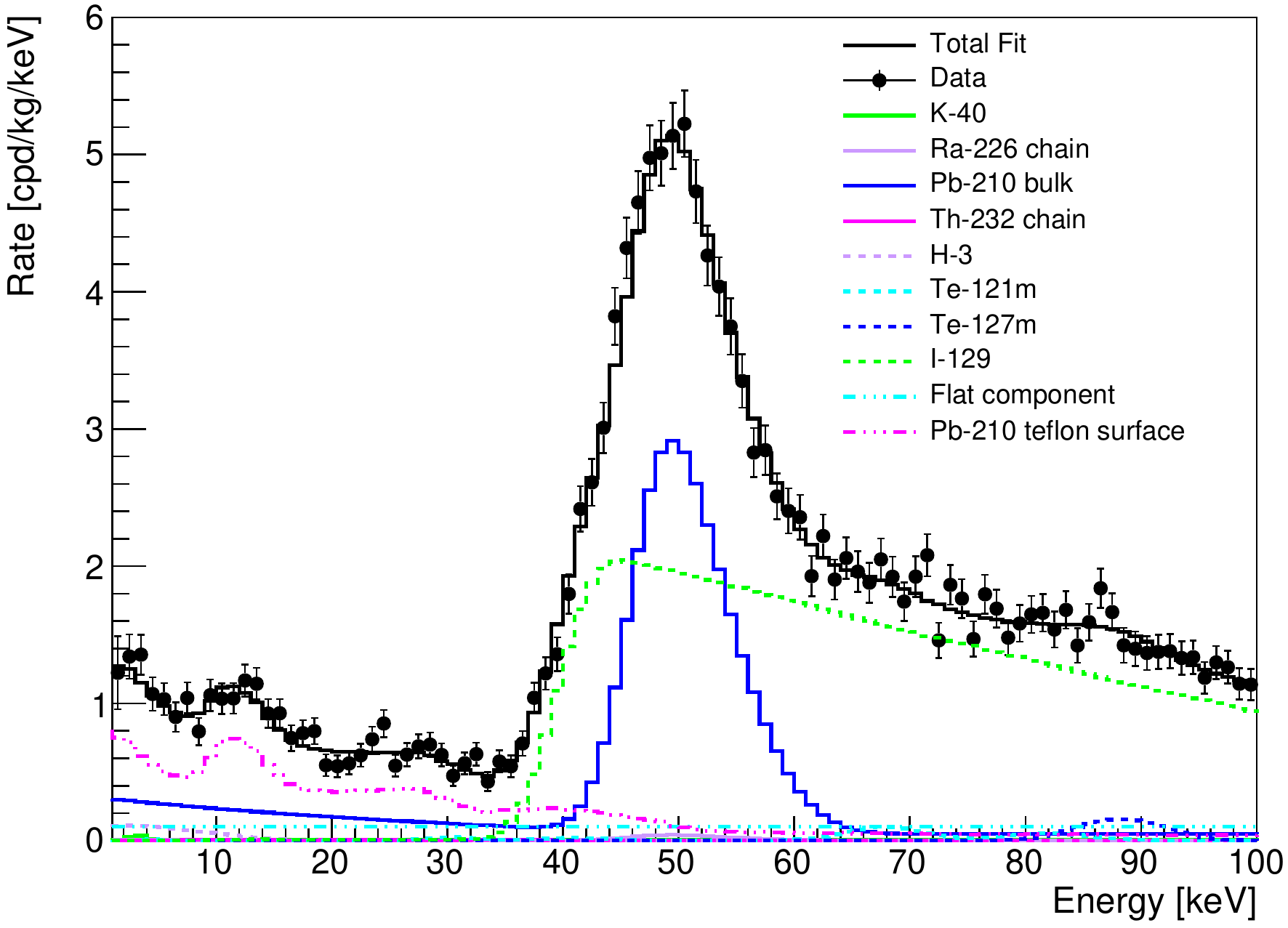}
  \caption{Energy spectrum from NaI-33 up to 100\,keV with a spectral fit. The spectrum is shown after selection cuts and efficiency correction.} \label{fig:3} 
\end{figure}

\begin{figure}[ht]
\includegraphics[width=\linewidth]{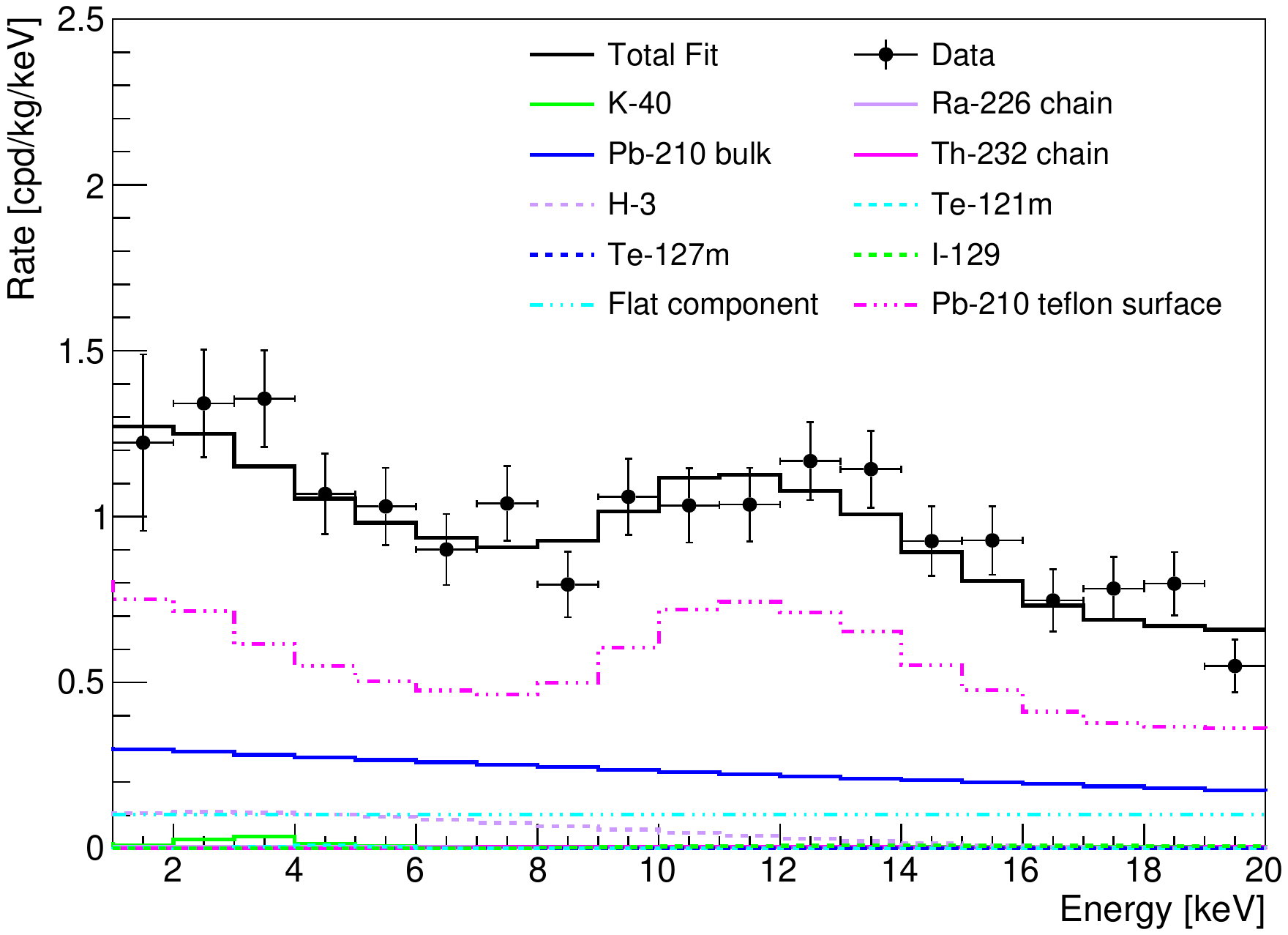}
\caption{Energy spectrum from NaI-33 below 20\,keV with the spectral fit. The spectrum is shown after selection cuts and efficiency-corrected.} \label{fig:4} 
\end{figure}

The light yield and FWHM resolution of NaI-33 crystal scintillator, measured by fitting the 59.5-keV photopeak of an $^{241}$Am source, is determined to be $12.1\pm0.2$~phe/keV and 13.5\%, respectively. The light yield is slightly higher than that measured in~\cite{nai033-udg}. %This can be attributed to nitrogen flushing of the inside of the crystal enclosure which has removed some moisture and improved the surface quality of the crystal.
To determine the background rate in the 1-6~keV ROI, data were taken between August 9, 2020 and September 5, 2020 for a total exposure of 26.4~days. To reject coincident backgrounds, events with energy larger than 50~keV in the veto are rejected with 42\% veto rejection power in the ROI. In addition, to reduce noise, the following cuts were applied:
\begin{itemize}
    \item $\langle t \rangle_{600} = \frac{\sum\limits_{t_i<600\textrm{ ns} }h_it_i}{\sum\limits_{t_i<600\textrm{ ns}}t_i} \in [140,270]$~ns;
    \item Trigger time delay~$\in [-36,36]$~ns;
    \item No. of clusters in each PMT~$\geq 2$;
    \item $C_{(0,1000)}/h_{max}$ $>$ 50\,ns;
    \item $0.2 < C_{(200,400)}/C_{(0,200)} < 0.9$;
    \item $0.1 < C_{(400,600)}/C_{(200,400)} < 0.9$.
\end{itemize}
where $h_i$ is the amplitude at time $t_i$ in ns, and $C_{(t_i,t_j)}$ is the pulse area between $t_i$ and $t_j$ in ns. These variables are described in detail in~\cite{nai033-udg}. Figure~\ref{fig:2} shows the final acceptance rate as a function of energy after these cuts. The average event acceptance in the ROI is estimated to be 77.6\%~\cite{phd-ambra}. The energy spectrum after cuts and efficiency correction is shown in Fig.~\ref{fig:3}~(black dots) up to 100~keV, and in Fig.~\ref{fig:4} up to 20~keV. The measured rate in 1-6~keV~(1-10~keV) is $1.20\pm0.05$~cpd/kg/keV~($1.09\pm0.04$~cpd/kg/keV).

%The final combined signal acceptance as a function of energy is shown in Fig.~\ref{fig:2}. 

%This result shows that, in terms of radiopurity, the NaI-33 is the best NaI(Tl) crystal ever produced after DAMA/LIBRA experiment, where the rate in 1-6~keV is of the order of 0.7~cpd/kg/keV~\cite{dama2}. 

% $\langle t \rangle_{600}$ \iffalse =\sum_{t_i<600}h_it_i / \sum_{t_i<600} h_i$, with $h_i$ being the pulse amplitude at time $t_i$ ;
% the Charge-over-Maximum, $CoM = C_{(0,1000)}/h_{max}$, where $C_{(t_i,t_j)}$ is the pulse area between $t_i$ and $t_j$

%The veto rejection power is equal to 42\%(47\%) in the energy window [1,6]([1,20]) keV.

A spectral analysis was performed to quantify the contribution of different background components~\cite{phd-ambra}. Predicted spectral shapes from different background sources were calculated by Monte Carlo~\cite{sabre-mc}. In the fit, the activities of the following components were treated as free or semi-free parameters: $^{40}$K, $^{210}$Pb, $^{3}$H, $^{226}$Ra, $^{232}$Th, $^{129}$I, $^{\mathrm{121m}}$Te, $^{\mathrm{127m}}$Te, and a flat component which includes $^{87}$Rb and other internal and external background contributions. In addition, $^{210}$Pb from the PTFE reflector wrapping the crystal was included to reproduce the peak at $\sim$12~keV due to X-rays from $^{210}$Pb. In the spectral fit, a Gaussian penalty was applied to $^{40}$K with mean equal to $0.14\pm0.01$~mBq/kg corresponding to the prediction from ICP-MS measurements~\cite{nai033-surface}. Assuming secular equilibrium for $^{226}$Ra and $^{232}$Th chain segments, Gaussian penalties~(\SI{5.9\pm0.6}{\micro\becquerel/\kg} for $^{226}$Ra and \SI{1.6\pm0.3}{\micro\becquerel/\kg} for $^{232}$Th) were implemented based on measurements from $^{214}$Bi-Po and $^{212}$Bi-Po time-correlated events, respectively~\cite{nai033-udg}.

The result of the fit is shown in Fig.~\ref{fig:3} and in Fig.~\ref{fig:4}. The p-value of the fit is equal to 0.26 with $\chi^2/N_{dof} = 96/88$. Table~\ref{tab:1} summarises the breakdown of the background components determined from the fit. The background rate in the ROI, dominated by $^{210}$Pb in the bulk of the crystal and on the surface of the PTFE reflector, is found to be conservatively equal to $1.16\pm0.10$ cpd/kg/keV. The $^{40}$K activity is estimated to be $0.14\pm0.01$~mBq/kg. This value is consistent with an independent measurement of $^{40}$K using the coincidence between 3.2-keV X-ray/auger electrons in the crystal and 1.46~MeV gamma ray in the active veto, which yielded $0.07\pm0.05$~mBq/kg~($2.2\pm1.5$~ppb, or $<$~4.7~ppb at 90\% CL) despite limited statistics. %This independent analysis also shows that the PoP veto detector can be used to measure K contamination with ppb-level sensitivity.

\begin{table}[htb!]
\caption{\label{tab:1}%
Background components in NaI-33 from the spectral fit, current rate in ROI~(1-6~keV), and projected rate in ROI for future crystals. The future rate assumes underground crystal production with zone-refining purification and improved reflector radiopurity. The activity of $^{210}$Pb in the reflector is normalized to the crystal mass for comparison with bulk activities. Upper limits are given as one-sided 90\% CL. Rates are conservatively calculated using upper limits. 
%Other backgrounds: a flat contribution that accommodates $^{87}$Rb and external $\gamma$-rays. The ROI corresponds to [1,6] keV.
}
\begin{ruledtabular}
\resizebox{\columnwidth}{!}{%
\begin{tabular}{lccc}
Source & Activity & Rate in ROI & Projected \\
       &  in NaI-33 &  in NaI-33 & rate in ROI \\
       &  [mBq/kg] & [cpd/kg/keV]   &  [cpd/kg/keV] \\
\hline
$^{40}$K & 0.14$\pm$0.01 & 0.018$\pm$0.001 & $\leq$0.004 \\
$^{210}$Pb (bulk) & 0.41$\pm$0.02 & 0.28$\pm$0.01 & $\leq$0.093$\pm$0.003 \\
\cline{2-4}
$^{226}$Ra & 0.0059$\pm$0.0006 & \multirow{2}{*}{0.0044$\pm$0.0005} & \multirow{2}{*}{0.0044$\pm$0.0005} \\
$^{232}$Th & 0.0016$\pm$0.0003 & & \\
\cline{2-4}
$^3$H & 0.012$\pm$0.007 & $\leq$0.12 & -- \\
\cline{2-4}
$^{129}$I & 1.34$\pm$0.04 & \multirow{3}{*}{$\leq$0.011} & \multirow{3}{*}{--} \\
$^{121\mathrm{m}}$Te & $\leq$0.084 & & \\
$^{127\mathrm{m}}$Te & 0.016$\pm$0.006 & & \\
\cline{2-4}
%\hline
%\bf{sub-total 1} & & $\leq$0.43$\pm$0.01 & $\leq$0.101$\pm$0.003 \\
%\hline
$^{210}$Pb (PTFE) & 0.32$\pm$0.06 & 0.63$\pm$0.09 & $\leq$0.007 \\ 
Other & & 0.10$\pm$0.05 & 0.10$\pm$0.05 \\
%\hline
%\bf{sub-total 2} & & 0.73$\pm$0.10 & $\leq$0.11$\pm$0.05 \\
\hline
\bf{total} & & 1.16$\pm$0.10 & 0.21$\pm$0.05 \\
\end{tabular}
}
\end{ruledtabular}
\end{table}

The activity of bulk $^{210}$Pb is determined to be $0.41\pm0.02$~mBq/kg, consistent with earlier measurements using $\alpha$ counting of $^{210}$Po~\cite{nai033-surface,nai033-udg}. Although this value is about one order of magnitude larger than that in DAMA, it is smaller than those in ANAIS and COSINE, where on average the activity is about 1~mBq/kg~\cite{bkg-anais, bkg-cosine}.

%, 
The activity of $^{210}$Pb in the reflector is measured to be $1.1\pm0.2$~mBq, with the first \SI{4}{\micro\meter} from the crystal surface mostly responsible for background events. Although the commercial PTFE tape was carefully cleaned by acid leach~\cite{phd-suerfu}, this analysis indicates that the cleaning procedure is not very effective in removing $^{210}$Pb impurities. The $^{210}$Pb radioactivity of PTFE has been studied by other experiments: the special PTFE used in the CUORE-0 experiment has a $^{210}$Pb activity \SI{\leq123}{\micro\becquerel\per kg_\textrm{PTFE}}~(90\% CL)~\cite{cuore0-bulk-u238,cuore-surface-pb}; in the DarkSide-50 experiment~\cite{ds50-1st-result}, $^{210}$Pb activity in PTFE is measured to be $\leq38$~mBq/kg$_{\textrm{PTFE}}$~(90\% CL) by $\gamma$-spectroscopy~\cite{teflon} and $\leq46$~mBq/kg$_{\textrm{PTFE}}$ by $^{210}$Po $\alpha$ counting~\cite{zuzel-po210}. These considerations imply that the background due to the PTFE reflector can be reduced to a secondary component by custom or special manufacturing of the PTFE tapes.

The activation rate of $^3$H in NaI is measured to be $83\pm27$~cpd/kg by ANAIS~\cite{anais-h3}. NaI-33 has undergone a 9-month surface exposure at sea level and a 12-month cooling underground. At the predicted activation rate, NaI-33 would have seen a $^3$H activity of \SI{37\pm12}{\micro\becquerel\per kg}. The actual $^3$H activity is estimated to be \SI{12\pm7}{\micro\becquerel\per kg}. Although limited by statistics, the $^3$H production rate in NaI-33 seems to be lower. One possibility is that the previously reported $^3$H rate is not entirely due to cosmogenic $^3$H, and instead varying amount of $^3$H is introduced into the crystal as NaOH impurity depending on the powder drying and crystal growth method~\cite{phd-suerfu}.

\ifPRL
\else
\section*{Future perspectives}
\fi

To further improve radiopurity of crystals to be grown in the future, we have tested the zone refining of ultra-high purity NaI powder. Table~\ref{tab:2}, reproduced from~\cite{zr-nai}, shows that many impurities are greatly reduced in the first three samples~(about 50\% of the ingot), and in particular $^{40}$K and $^{87}$Rb are reduced to negligible levels.

\begin{table}[htb!]
\caption{\label{tab:2}%
In the zone refining test, 53 zone passes were performed on 744~g of ultra-high purity NaI powder, and five samples were taken from different ingot locations and analyzed to determine the purification efficiency~\cite{zr-nai}. A few other elements are included to show effectiveness for other metallic impurities.}
\begin{ruledtabular}
\begin{tabular}{lcccccc}
Isotope & \multicolumn{6}{c}{Impurity concentration~(ppb)} \\
\cline{2-7}
        & Powder & $S_1$ & $S_2$ & $S_3$ & $S_4$ & $S_5$ \\
\hline
$^{39}$K & 7.5 & $<$0.8 &  $<$0.8 & 1 & 16 & 460 \\
$^{85}$Rb & $<$0.2 & $<$0.2 & $<$0.2 & $<$0.2 & $<$0.2 & 0.7 \\
$^{208}$Pb & 1.0 & 0.4 & 0.4 & $<$0.4 & 0.5 & 0.5 \\
$^{24}$Mg & 14 & 10 & 8 & 6 & 7 & 140 \\
$^{133}$Cs & 44 & 0.3 & 0.2 & 0.5 & 3.3 & 760 \\
$^{138}$Ba & 9 & 0.1 & 0.2 & 1.4 & 19 & 330 \\
\end{tabular}
\end{ruledtabular}
\end{table}

Quantitatively, the reduction of K and Rb depends on the segregation coefficient~$k$ and the fraction of the purified material to be reserved~\cite{phd-suerfu,zr-nai}. The segregation coefficient of K is estimated to be 0.57 while for Rb it is $<$~0.59 at 90\% CL~\cite{zr-nai}. Based on this, Table~\ref{tab:3} lists the reduction of impurity concentration for different combinations of number of zone passes and fraction of the material reserved~(Fig.~9 in~\cite{zr-nai}). Based on this, the purity can be further improved by a factor of~10 with 25~zone passes~(1-week processing time) and at the cost of 20\% of the initial material, or at the same cost of material but with 50~zone passes~(2-week processing time), the average purity can be improved by a factor of~25. Therefore, for future NaI crystals, the background due to $^{40}$K and $^{87}$Rb can be made subdominant compared to the amplitude of the modulation.

\begin{table}[ht!]
\caption{\label{tab:3}
Reduction of K for different numbers of passes and when different fractions of the ingot is reserved.}
%\begin{ruledtabular}
\begin{tabular}{cccccc}
\hline
\hline
\multirow{2}{*}{No. of passes} & \multicolumn{5}{c}{Impurity reduction}\\
\cline{2-6}
& \multicolumn{5}{c}{Fraction of material retained} \\
        & 50\% & 60\% & 70\% & 80\% & 90\% \\
\hline
10 & 0.3 & 0.36 & 0.42 & 0.48 & 0.56\\
25 & 0.03 & 0.050 & 0.07 & 0.11 & 0.25\\
50 & 0.001 & 0.0026 & 0.0086 & 0.037 & 0.18\\
\hline
\hline
\end{tabular}
%\end{ruledtabular}
\end{table}

Pb is more or less uniformly reduced by a factor of $\approx$3 without following the typical impurity distribution of zone refining~\cite{zr-nai}. Although one can still exploit this factor of~3, zone refining is not as efficient in removing $^{210}$Pb. Thus we have started new R\&D activity to explore alternative purification methods and have achieved preliminary progress. Since the analysis of data on the removal of Pb is still in progress, in this article, we adopt the conservative reduction factor for $^{210}$Pb obtained from zone refining alone.

%{\color{red} This result is consistent with the hypothesis that gaseous lead components distribute uniformly within the tube, originating a constant reduction factor in contrast to the other components which tend to accumulate in one direction. Based on this finding we are working on a method to further reduce $^{210}$Pb contamination. In particular, we plan to purge with a gas flow in the same direction as the oven moves along the tube during ZR.  This will move lead and its compounds, PbI$_2$ and PbCl$_2$, which are volatile at the NaI melting temperature,
%This will move all the Pb (MP = 327\,\textdegree{}C) and its compounds, PbI$_2$ (MP = 402\,\textdegree{}C)  and PbCl$_2$ (MP = 501\,\textdegree{}C) to the far end of the tube. }
%In principle, this procedure could lead to a higher reduction factor than currently observed.
%In principle, this procedure should remove most of the Pb and its compounds. Therefore, we expect a reduction factor for $^{210}$Pb greater than what shown in Table \ref{tab:2}, which is about a factor of three. 

With the aforementioned measures, a projected background rate of 0.21~cpd/kg/keV can be practically achieved in the future SABRE crystals~(see Table~\ref{tab:1}). In this projection, we have assumed the contamination level of NaI-33 for $^{40}$K and bulk $^{210}$Pb scaled by the reduction factor of zone refining in Table~\ref{tab:3}. For $^{210}$Pb in the reflector, we have assumed a contamination equal to the upper limit for the PTFE measured in DarkSide-50. To suppress cosmogenic backgrounds to a negligible level, we are also investigating the possibility of establishing an underground crystal growth facility at the Canfranc Laboratory in Spain~\cite{lab-canfranc}.

%Yet, we conservatively neglect this and include as an upper limit for $^{210}$Pb the contamination scaled for the present reduction factor.
%Yet, in this work we conservatively assume an upper limit for $^{210}$Pb after ZR by reducing our present measurement in NaI-33 by only a factor of three and quoting an upper limit, as reported in Table \ref{tab:1}.
%In addition, we reduce $^{40}$K according to measured reduction factors in Table \ref{tab:2} for $^{39}$K.
%In addition, we reduce $^{40}$K and $^{87}$Rb according to measured reduction factors in Table \ref{tab:2} for $^{39}$K and $^{85}$Rb.
%As far as $^{210}$Pb from the reflector is concerned, we assume a conservative contamination level equal to the one observed in PTFE for DS-50 \cite{bib:teflon, bib:ds50}. With these considerations the projected rate in the ROI is $\leq$\,0.21 cpd/kg/keV. 
%This shows that a background of order $<$0.2 cpd/kg/keV seems feasible once a high radiopurity reflector is produced and ZR purification exploited. Ultimately, we plan to grow crystals underground, which sensibly reduces cosmogenic activation of NaI.
%we notice that underground growth will reduce cosmogenic components.
%Therefore, we assume underground growth in our projected rate in Table \ref{tab:1}. The possibility to make an underground facility at the Canfranc Laboratory in Spain \cite{bib:lsc} is under investigation.

To quantify the sensitivity to the dark matter-induced annual modulation, the figure-of-merit~(FoM) defined in~\cite{anais-stat} can be used where a small FoM indicates a high sensitivity. In this framework, the DAMA/LIBRA phase II~\cite{dama2} has FoM $ = 8\times 10^{-4}$~d$^{-1}$kg$^{-1}$keV$^{-1}$. A similar value of FoM can be obtained with a 60~kg$\times$5~yr exposure and the predicted background rate of 0.21~cpd/kg/keV in Table~\ref{tab:1}. With the same exposure and background rate, the minimum detectable rate~\cite{msr-det-radiation} is expected to be $\leq(2\pm1)\times10^{-3}$~cpd/kg/keV at 90\% CL, which is about 5 times less than DAMA modulation amplitude of $10.5\pm1.1\times10^{-3}$~cpd/kg/keV in 1-6~keV ROI~\cite{dama2}. On the other hand, for the current generation of NaI(Tl) detector arrays, several decades are needed to obtain the same FoM as DAMA/LIBRA phase II. Therefore, the present results obtained from NaI-33 together with the planned improvements discussed above will be fundamental for the next generation NaI-based dark matter detectors.
%~(total mass on the order of 100~kg and background rate about 3-5~cpd/kg/keV in 1-6~keV ROI)
\iffalse
\begin{equation}
    FoM = \sqrt{\frac{2 \cdot b}{M \cdot t \cdot \Delta E}}
\end{equation}
where $b$ is the background rate in the ROI with width $\Delta E$. $M$ is the detector mass, and $t$ is the exposure time. A small FoM value indicates a high sensitivity to annual modulation. In particular, for DAMA/LIBRA phase II \cite{bib:dama2} $b \sim 0.7$\,cpd/kg/keV in [1,6]\,keV for an exposure equal to 1.13 ton$\times$yr, which gives $FoM = 8\times 10^{-4}$\,cpd/kg/keV.
\fi

%We have also determined the minimum detectable rate \cite{bib:ld} for 5 years data taking and a total mass of 60\,kg to be $\leq(2\pm1)\times10^{-3}$\,cpd/kg/keV at 90\% C.L. This is less than the DAMA modulation amplitude, $S_m = 0.0105\pm0.0011$\,cpd/kg/keV in [1,6]\,keV \cite{bib:dama2}.
%Therefore, the impact of the present result is crucial for the development of future NaI-based detectors to probe DM induced annual modulation.

\ifPRL
\else
\section*{Conclusions}
\fi

%Thanks to the excellent performance of the SABRE PoP setup with the liquid scintillator active veto, K concentration in  is confirmed to be below 4.7~ppb in this crystal. Furthermore, 

The growth of NaI-33 and its characterization in the SABRE detector are a breakthrough in the technology to support next-generation NaI experiments to probe the long-standing DAMA result: the background rate in the 1-6~keV ROI is determined to be $1.20\pm0.05$~cpd/kg/keV, several times lower than all presently running NaI-based experiments except DAMA. At present, the dominant backgrounds are due to bulk $^{210}$Pb and $^{210}$Pb in the PTFE reflector. With a combination of zone refining, special low-radioactivity PTFE tape, and underground growth of crystals, the background in the ROI in future crystals can be further reduced to $\leq0.21$~cpd/kV/keV. This result marks a significant improvement in the quest for testing the dark matter-induced annual modulation, and paves the road to the next-generation NaI-based experiments with higher sensitivities.

%We have shown that NaI-33 background can be significantly improved beyond the present radiopurity of the DAMA crystals, a result still unsurpassed.

\ifWordCount
\else
\begin{acknowledgments}
\textit{
This work has been supported by INFN funding and National Science Foundation under the award number PHY-1242625, PHY-1506397 and PHY-1620085. We thank Ezio Previtali, Andrea Giuliani, and Monica Sisti for discussions about PTFE radiopurity in CUORE. We thank Fausto Ortica and Aldo Romani from Perugia University for support in photospectrometric measurements. We thank the Gran Sasso Laboratory for the support during the installation of the SABRE PoP setup. }
%We thank M. Laubenstein and P. Meyer for sharing with us measurements on teflon radiopurity in DarkSide-50.
\end{acknowledgments}
\fi

\ifWordCount
\end{document}
\fi

\bibliographystyle{apsrev4-1}
\bibliography{reference.bib}

\end{document}